\documentclass[doublecol]{epl2}  

\RequirePackage{subfigure}
\RequirePackage{amsmath}    
\RequirePackage{graphicx}

\RequirePackage[a4paper, total={6in, 8in}]{geometry}

\RequirePackage{verbatim}  
\RequirePackage{color}      

\title{Exploring effects of magnetic field on the Hadron Resonance Gas}

\author{Abhijit Bhattacharyya\inst{1}
        \and
        Sanjay K. Ghosh \inst{2}
        \and
        Rajarshi Ray \inst{2}
        \and
        Subhasis Samanta \inst{2}
}

\institute {
\inst{1}{Department of Physics, University of Calcutta, 92, A. P. C. Road, Kolkata - 700009, India}
           
\inst{2}{Center for Astroparticle Physics \& Space Science, Bose Institute, Block-EN, Sector-V, Salt Lake, 
           Kolkata-700091, India}

}

\pacs{25.75.-q}{Relativistic heavy-ion collisions}
\pacs{12.38.Mh}{Quark gluon plasma}

\abstract{
We present a study of the effects of magnetic fields on fluctuations and
correlations in hadron resonance gas model. We find significant  changes
in the fluctuations of net baryon number, electric charge and
strangeness. This is also reflected in various fluctuation ratios along
the freezeout curve.
}

\begin{document}
\maketitle
\section{\label{sec:Intro} Introduction}

Heavy-ion collisions (HIC) are investigated both theoretically and
experimentally to understand the properties of nuclear matter at extreme
conditions.  One of the most important issues  addressed in HIC is the
possibility for nuclear matter to undergo a phase transitions to quark
matter. At low baryon density and high temperature nuclear matter is
expected to smoothly cross over~\cite{Aoki_Nature} to a quark gluon
plasma (QGP) phase. Whereas, at high baryon density and low temperature
the system is expected to have a first order phase
transition~\cite{Asakawa_NPA_504,Ejiri_PRD_78,Bowman_PRC_79}.

The study of the effect of magnetic field on the phase transition has
become a subject of intense research for last few years. The phase
transition in Quantum Chromodynamic (QCD) system is usually expected to
occur around the QCD energy scales $\Lambda \sim$ 200 MeV. So one should
be interested in the magnetic fields with strength B $\sim$ (200
MeV)$^2$ $\sim 2\times 10^{18}$~G.  Non-central relativistic HIC may
create extremely strong magnetic field ($\sim m_{\pi}^2 \sim 10^{18} G$)
due to the relativistic motion of the charged particles. The 
magnetic field ($B$) may reach up to order 
of  $0.1 m_{\pi}^2$, $m_{\pi}^2$ and $15 m_{\pi}^2$ for SPS, RHIC and LHC energies
respectively~\cite{Skokov}. 

Magnetic fields can induce many interesting phenomena in QCD matter.
For example the chiral magnetic effect, i.e.,  electric charge separation
induced by chirality imbalance, along an external magnetic field, which
also results in P and CP violation~\cite{schafer,Fukushima_PRD78}.  On
the other hand, magnetic catalysis~\cite{Shovkovy_2013} and inverse
magnetic catalysis~\cite{Preis_2011,Bruckmann_2013} can affect the phase
diagram of QCD matter. 

Lattice QCD studies show that in the presence of magnetic field the
critical temperature may increase \cite{Swagato_PRD_82,Bali_JHEP_02}.
Effects on hadron mass modification has also been reported~\cite{latmass}.
The presence of magnetic fields may increase the fluctuations and
correlations~\cite{Fu_PRD_88}, as well as elliptic flow 
coefficient~\cite{Srivastava_MPL_A26} of hot QCD matter.  Anisotropic electric
conductivity of hadronic matter is an effect of strong magnetic field
which has been found in lattice studies~\cite{Buividovich_prl2010}. Such
anisotropy in conductivity should create an anisotropy in the dilepton
emission rate with respect to reaction plane and should be 
observed~\cite{Bratkovskaya_1995,Gupta_2004}.  The final observables
in relativistic HIC may also depend on the evolution of the magnetic
field.  Depending on the electrical conductivity of the medium the
magnetic diffusion time $\tau_{mag}$ may vary from 0.3 fm to 150
fm~\cite{Tuchin_PRC_82,Deng_2012}. So for large diffusion time a constant magnetic field 
may be considered.

In view of the importance of magnetic fields, we plan to study its
effect on fluctuations and correlations of strongly interacting matter,
using the Hadron Resonance Gas (HRG)
model~\cite{PLB344_Braun-Munzinge,arXiv:nucl-th/9603004_Cleymans,Cleymans_PRC_73,PLB695_Karsch,
Endrodi_HRG,SSamant_PRC_90,SSamanta_PRC_91_041901}.  Here the confined
phase of QCD is modelled as a non-interacting gas of hadrons and
resonances. Recently, the hadronic EOS for non-zero magnetic fields has
been studied within the HRG model~\cite{Endrodi_HRG}. On the other hand,
study of fluctuations and correlations of conserved charges is a
reliable way to study the phase transition. These quantities behave
quite differently in the hadronic and quark
phases~\cite{Jeon_PRL_85,Asakawa_PRL_85,Ejiri_PLB_633}.  Moreover, the
fluctuations are expected to be enhanced near phase boundary and are
related to the critical behaviour of strongly interacting
matter~\cite{Stephanov_PRL_81,Hatta_PRD_67}. Furthermore, by studying
event-by-event fluctuations~\cite{Stephanov_PRD_60} as a function of
beam energy~\cite{STAR_PRL_112,STAR_arXiv:1402.1558} it would be
possible to learn more about the QCD phase diagram.  The paper is
organised as follows. First we discuss the HRG model in the presence of a
magnetic field. Thereafter we present our results and finally we
conclude.

\section{HRG model in presence of magnetic field}

The grand canonical partition function of HRG may be written as sum of
all the partition functions $Z_i$ of each hadron $i$, where
\begin {align}
\ln Z_i 
=\pm Vg_i \int \frac{d^3p}{{(2\pi)}^3} \ln[1\pm e^{-(E_i-\mu_i)/T}].
\end{align}
\noindent
Here $V$ is the volume of the system, $g_i$ is the degeneracy factor,
$T$ is the temperature, $E_i=\sqrt{{p}^2+m^2_i}$ is the single-particle
energy, $m_i$ is the mass and $\mu_i=B_i\mu_B+S_i\mu_S+Q_i\mu_Q$ is the
chemical potential. In the last expression, $B_i$, $S_i$, $Q_i$, are
respectively, the baryon number, strangeness and charge of the particle,
$\mu^,$s are corresponding chemical potentials. The upper and lower
signs corresponds to fermions and bosons respectively.
We have incorporated all the hadrons listed in the particle
data book~\cite{PDG2014} up to mass of 3 GeV.

\begin{figure*}[!htb]
\centering
  \subfigure[]{\includegraphics[scale=0.35]{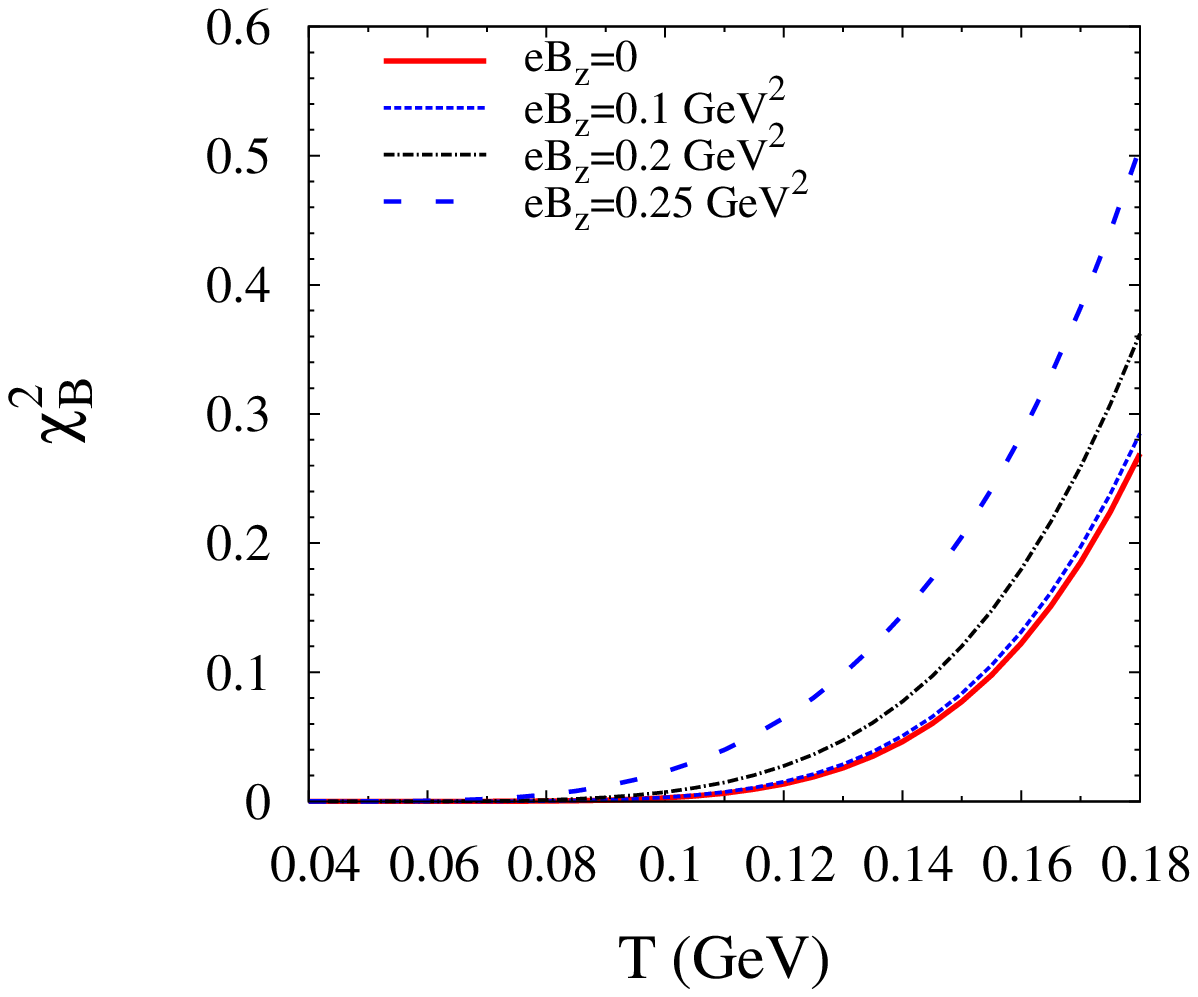}\label{fig:chi_B2-T}}
  \subfigure[]{\includegraphics[scale=0.35]{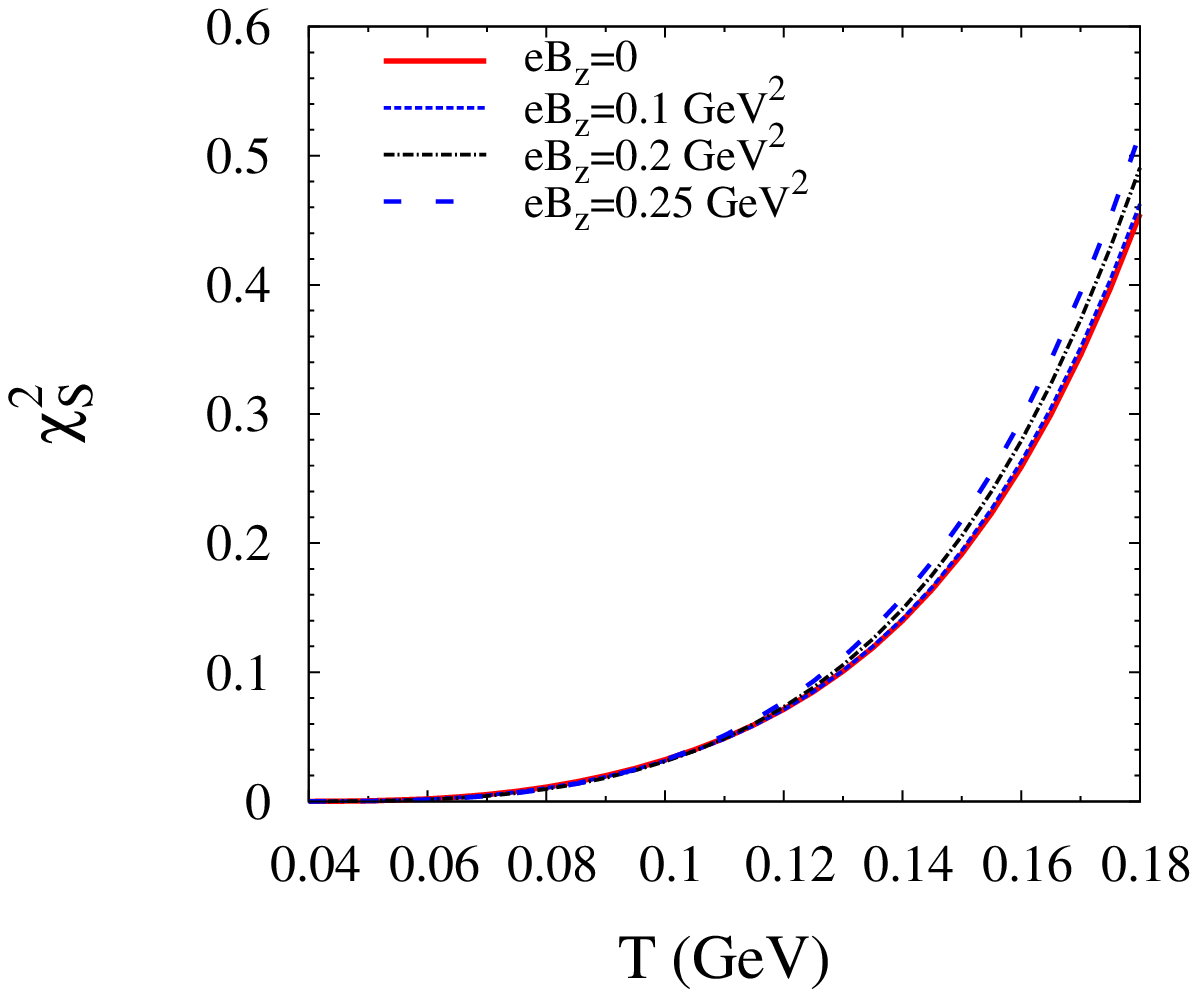}\label{fig:chi_S2-T}}
  \subfigure[]{\includegraphics[scale=0.35]{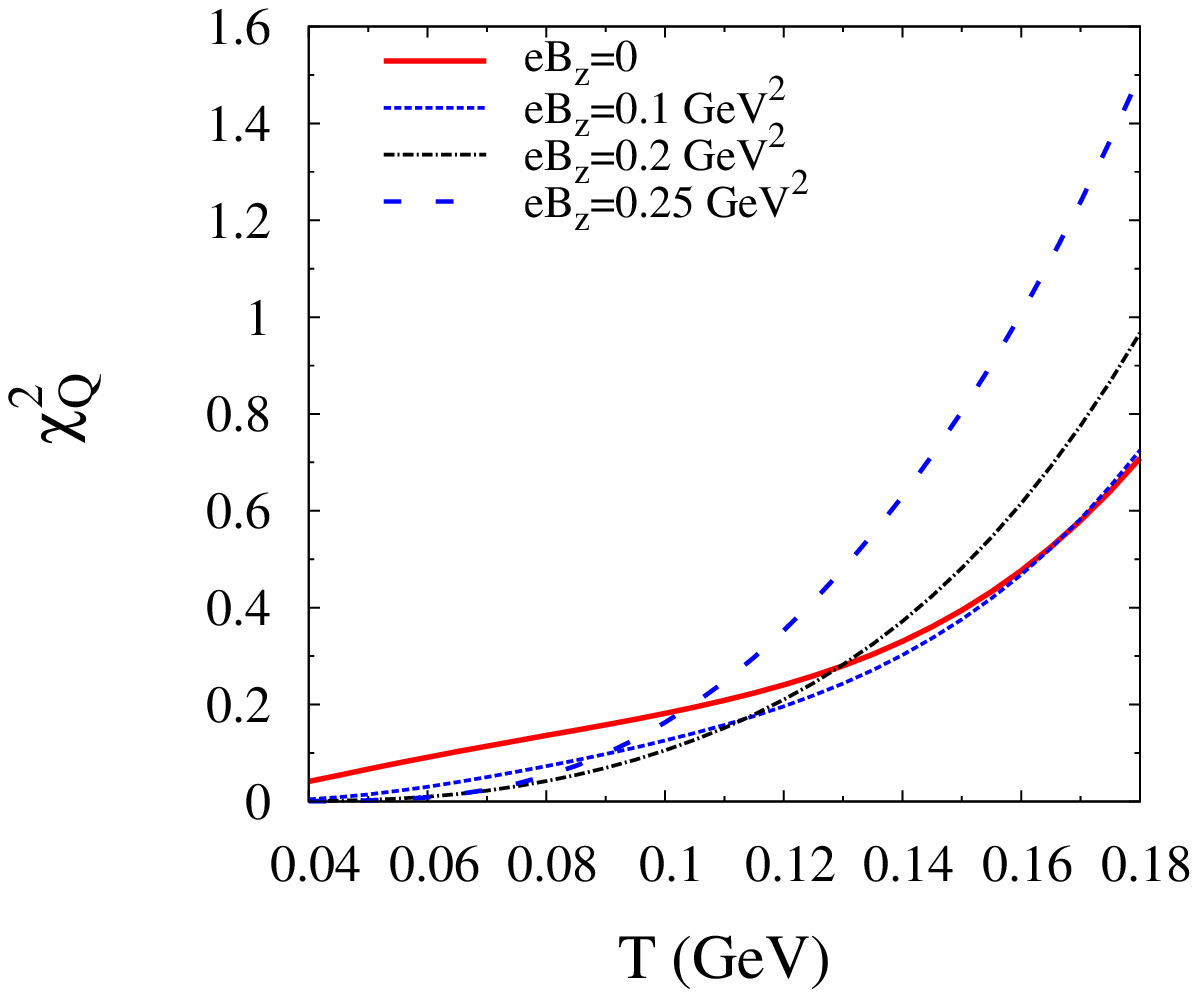}\label{fig:chi_Q2-T}}
\subfigure[]{\includegraphics[scale=0.35]{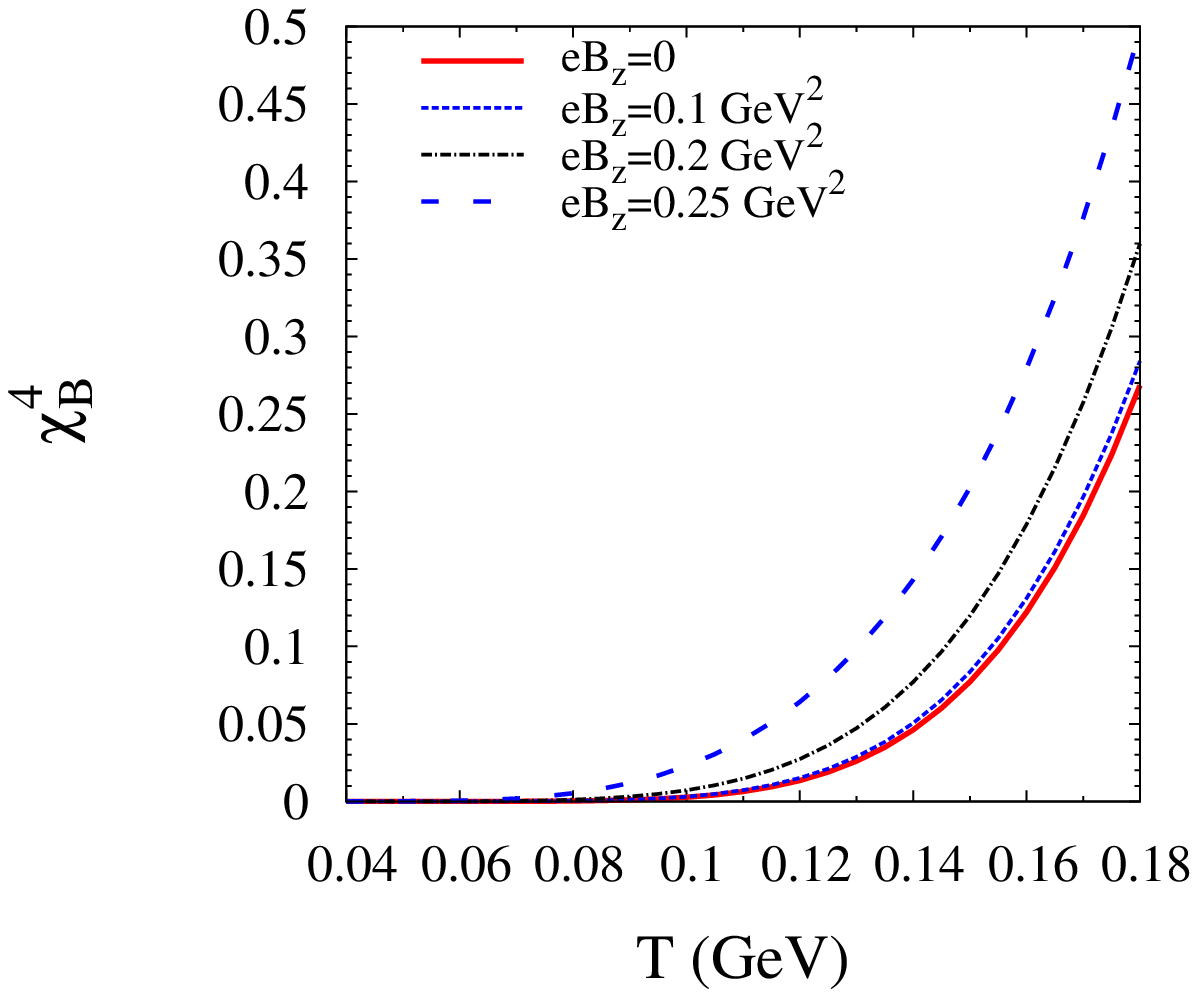}\label{fig:chi_B4-T}}  
\subfigure[]{\includegraphics[scale=0.35]{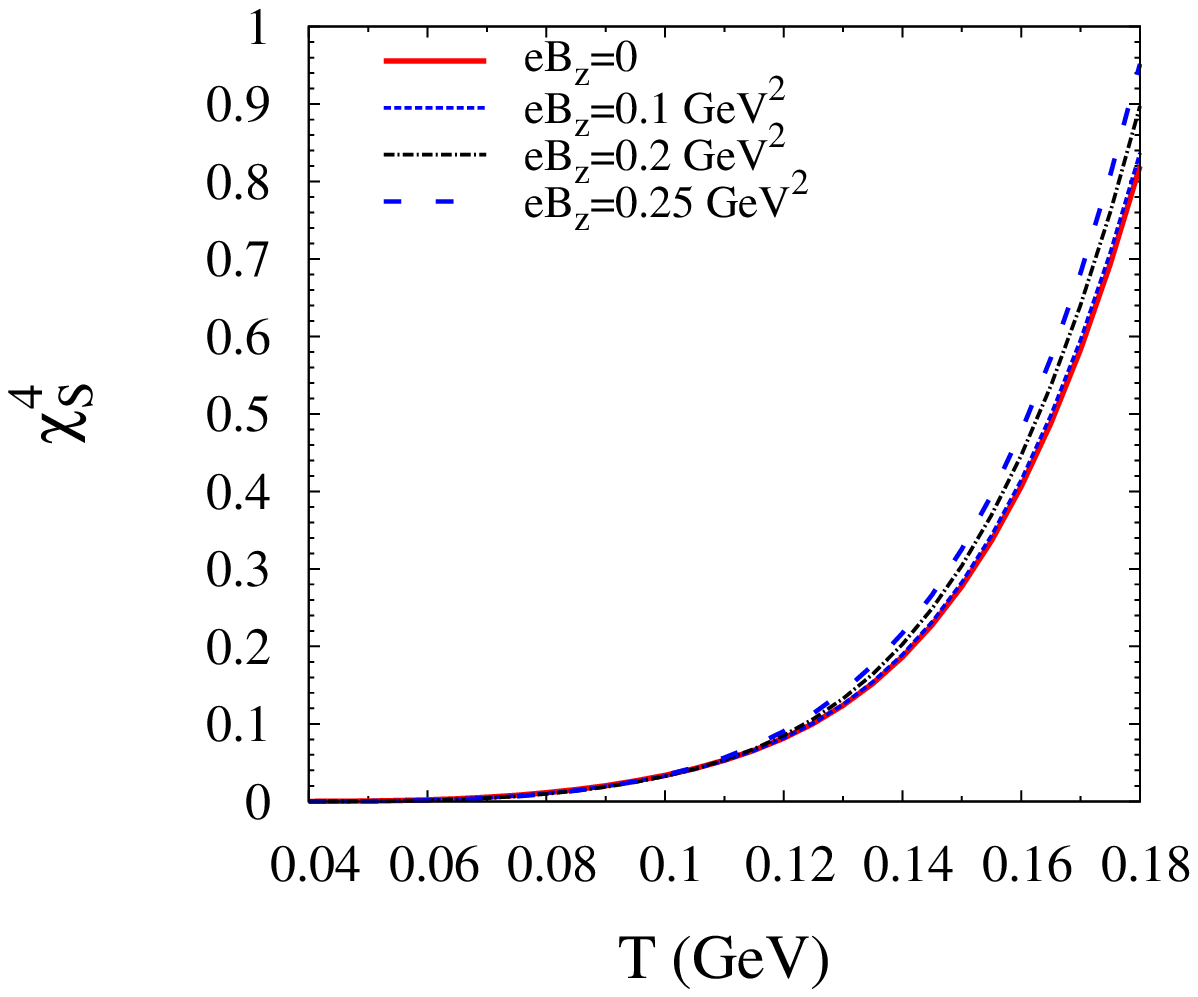}\label{fig:chi_S4-T}}  
\subfigure[]{\includegraphics[scale=0.35]{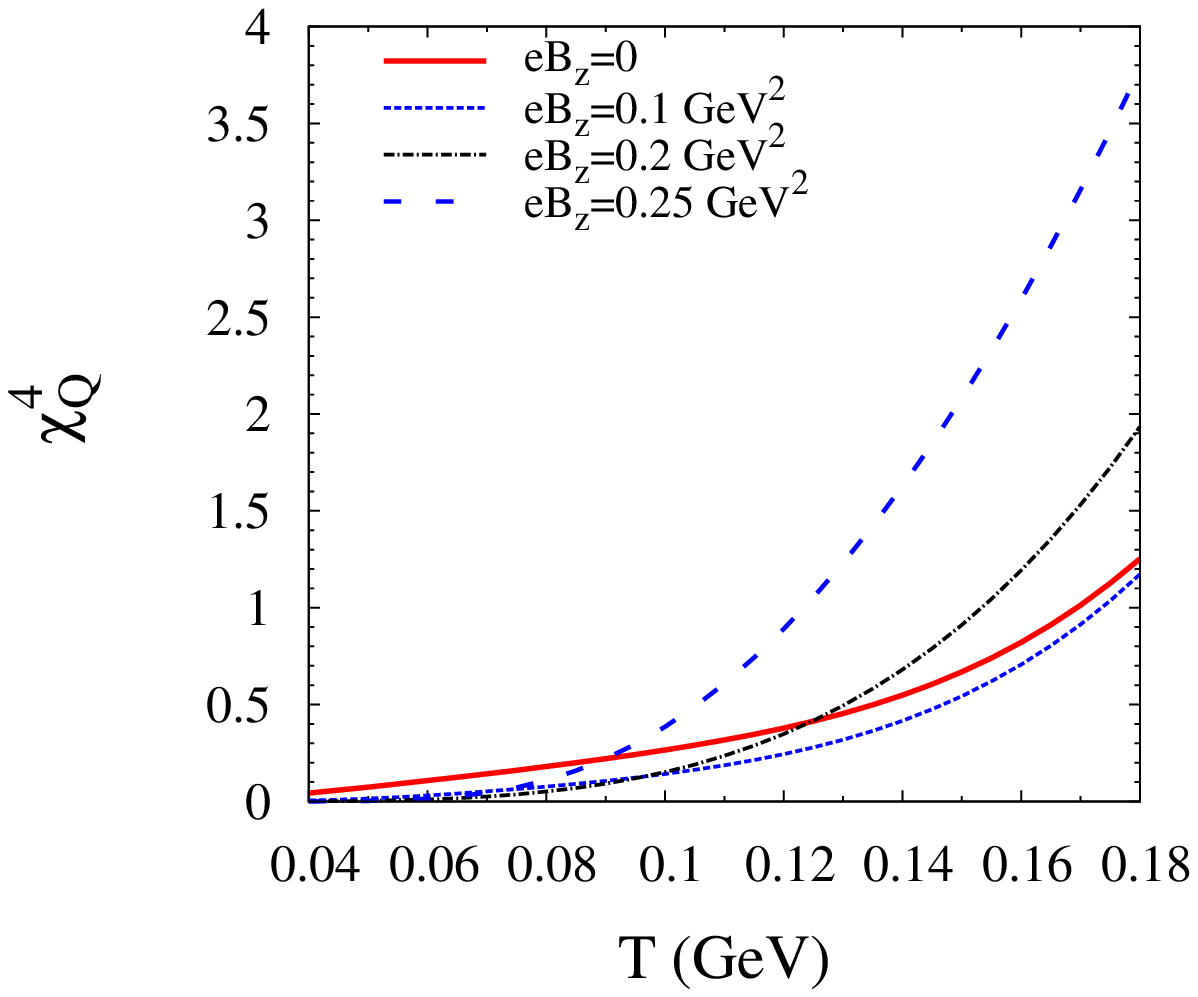}\label{fig:chi_Q4-T}}  
 \caption{Variation of $\chi^2$ and $\chi^4$  with $T$ at $\mu=0$ for baryon number, 
strangeness  and electric charge.}
 \label{fig:chi_T}
\end{figure*} 

For a constant magnetic field $B_z$, the single particle
energy levels for charged and neutral particles are respectively
given by~\cite{Bjorken,Broderick:2000pe,Broderick:2001qw}

\begin{align}
&E_{i,c}^{B_z}=\nonumber \\
&\sqrt{p_z^2+ (\sqrt{ m_i^2+|q_i| B_z(2n+2s_z+1)}+2s_z k_i B_z)^2}\\
&{\rm and} \nonumber \\
&E_{i,n}^{B_z}=\sqrt{p_z^2+ \left(\sqrt{m_i^2+p_x^2+p_y^2}+
2s_z k_i B_z\right)^2}, 
\end{align}

\noindent
where $q_i$ is the charge of the particle $n$ is any positive integer
corresponding to allowed Landau levels, $s_z$ are the components of spin
$s$ in the direction of magnetic field, and $k_i$ is the anomalous
magnetic moment.
For a given $s$, there are $2s+1$ possible values of $s_z$. The
gyromagnetic ratios are taken as $g_h=2|q_h/e|$ for all charged hadrons~\cite{Ferrara,Endrodi_HRG}.  
Therefore the pressure for $i^{th}$ hadron
may be written as

\begin{align}
 P_i =\frac{g'_iT}{{(2\pi)}^3} \sum_{s_z} \int \pm d^3p 
  & \ln[1\pm e^{-(E_{i,n}^{B_z}-\mu_i)/T}], \nonumber \\ 
  & ~~~for~ Q_i=0, 
\end{align} 
\begin{align}
 P_i &=\frac{g'_i T}{2\pi^2}|Q_i| e B_z \sum_{n} \sum_{s_z} \int_{0}^{\infty}
 \pm dp_z \nonumber \\
 &  \ln[1\pm e^{-(E_{i,c}^{B_z}-\mu_i)/T}], ~~~for~ Q_i\neq 0
\end{align} 
where $g'$ is the degeneracy other than spin.
There are several
simplifying assumptions that we are using here. Firstly we are
considering a non-interacting HRG. Secondly we are considering
point-like hadrons with their masses remaining unaffected by the ambient
temperature, chemical potentials and magnetic fields. These are common
assumptions that are regularly used to study the strongly interacting
matter in the context of heavy-ion collisions as may be found in the
Refs.~\cite{PLB344_Braun-Munzinge,
arXiv:nucl-th/9603004_Cleymans,Cleymans_PRC_73,PLB695_Karsch,
Endrodi_HRG,SSamant_PRC_90,SSamanta_PRC_91_041901} and references
therein. Similar assumptions are also used by various lattice QCD groups
to set the scales of their results for various thermodynamic quantities~\cite{Bazavov:2012jq,Borsanyi:2010cj}. 
This is true also in the studies
of magnetised neutron stars as may be found in
Refs.~\cite{Chakrabarty:1997ef,Bandyopadhyay:1997kh,Broderick:2000pe,
Broderick:2001qw,Mallick:2014faa}.

Also, apart from the above thermal contribution, there is also a vacuum part
which is in general divergent and needs regularisation and
renormalisation. However, theories with particles having spin $>1$ are
not renormalizable~\cite{Peskin:1995ev}.  This is probably reflected
through the fact that the renormalized contributions from particles with
spin $> 1$ in the HRG vacuum is negative~\cite{Endrodi_HRG}. Since many
of the thermodynamic quantities including fluctuations and correlations
of conserved charges are unaffected by the vacuum part, we have
neglected its contribution altogether.

\section{Results}\label{secresults}

The $n^{th}$ order fluctuation and correlation are given respectively by
the diagonal and off-diagonal components of the susceptibility

\begin{equation}\label{eq:chi}
 \chi^{jk}_{xy}=\frac{\partial^{j+k} {(\sum_{i} P_i/T^4)}}
{\partial {(\frac{\mu_x}{T})}^j  
\partial {(\frac{\mu_x}{T})}^k}, 
\end{equation}
where $n=j+k$, $\mu_{x/y}$ is the chemical potential for conserved charge
$x/y$, with $x,y=B~/~S~/~Q$. 

\begin{figure*}[!htb]
\centering
\subfigure[]{\includegraphics[scale=0.39]{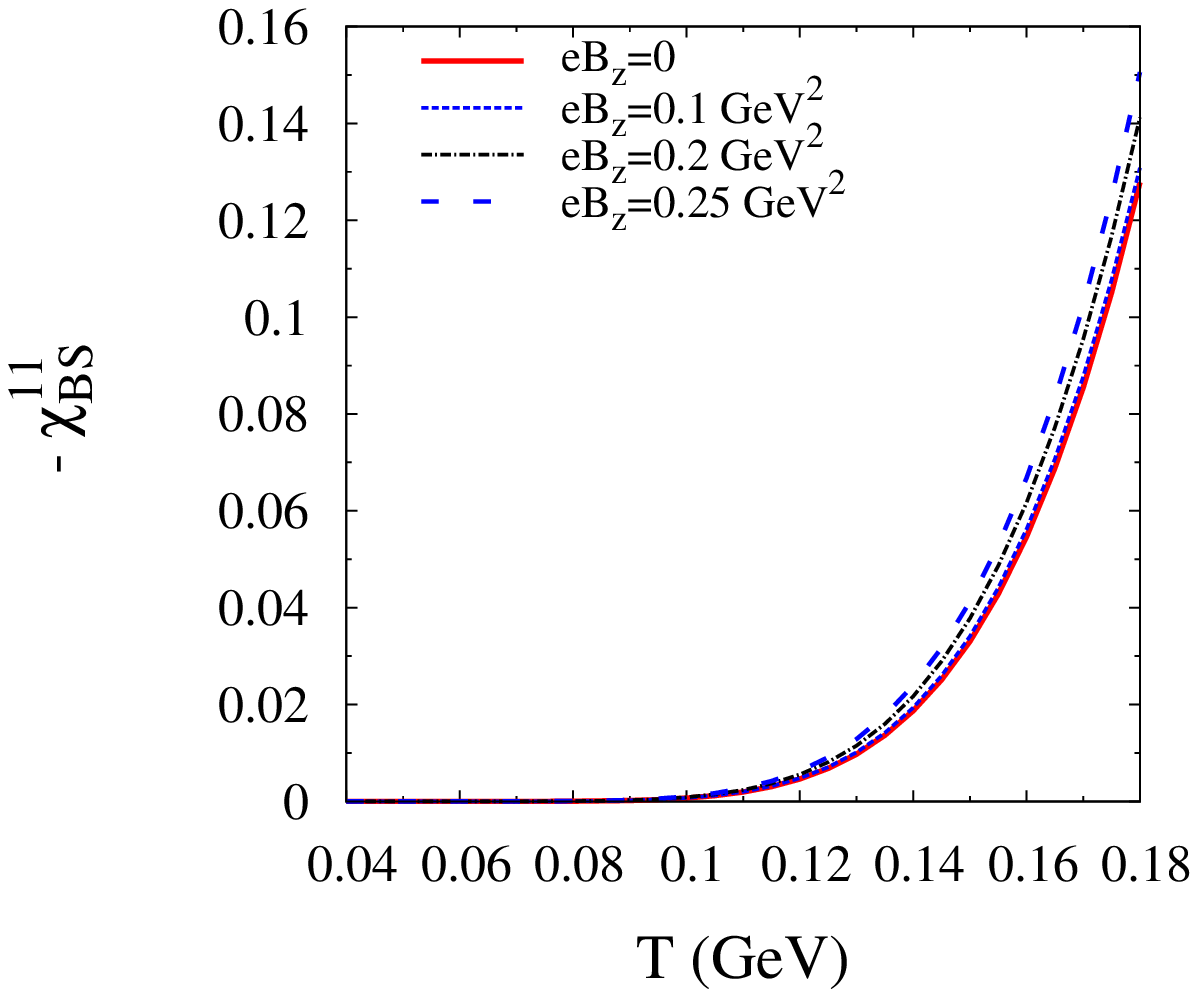}\label{fig:chi_BS_T}}
\subfigure[]{\includegraphics[scale=0.39]{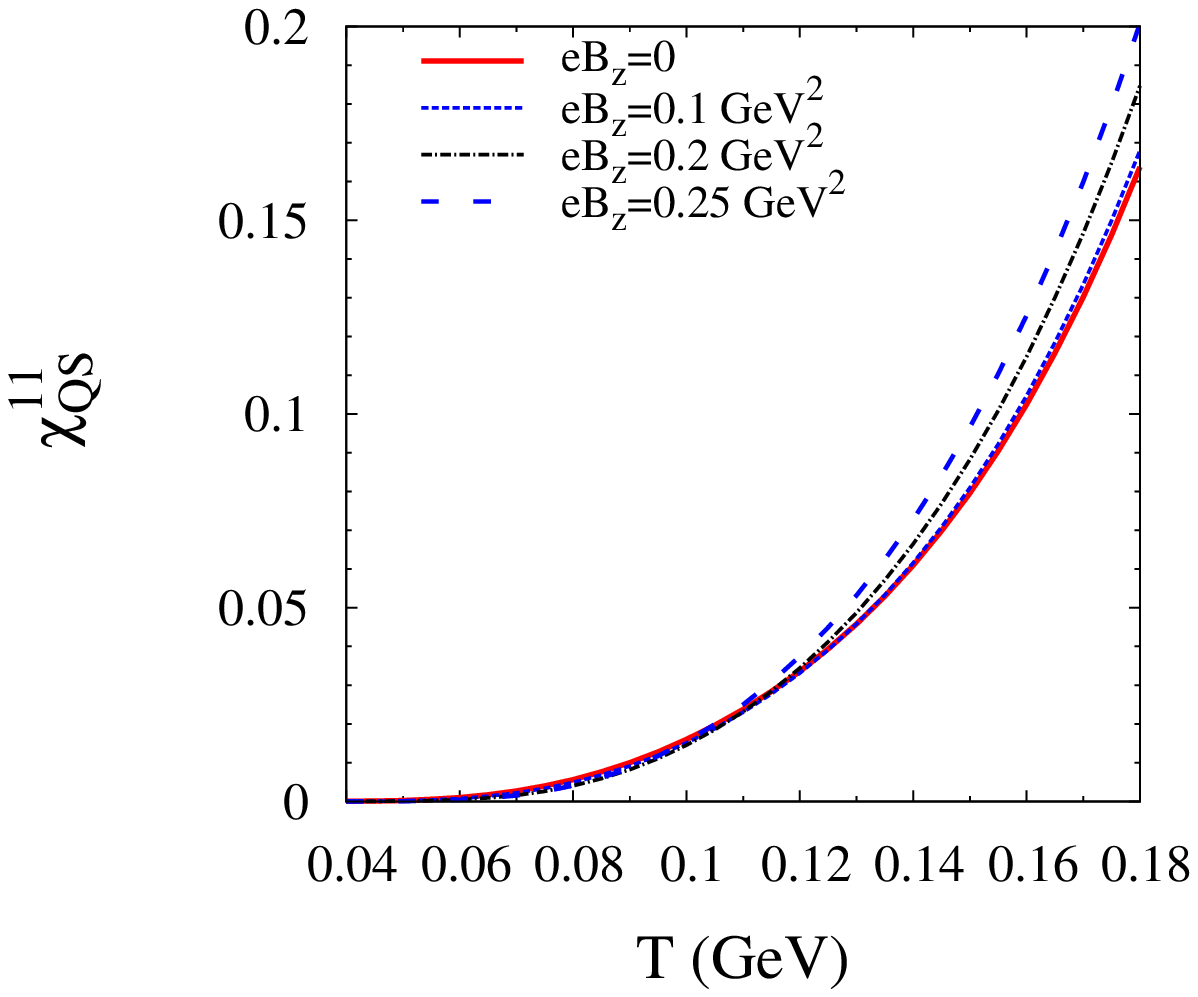}\label{fig:chi_QS_T}}
\subfigure[]{\includegraphics[scale=0.39]{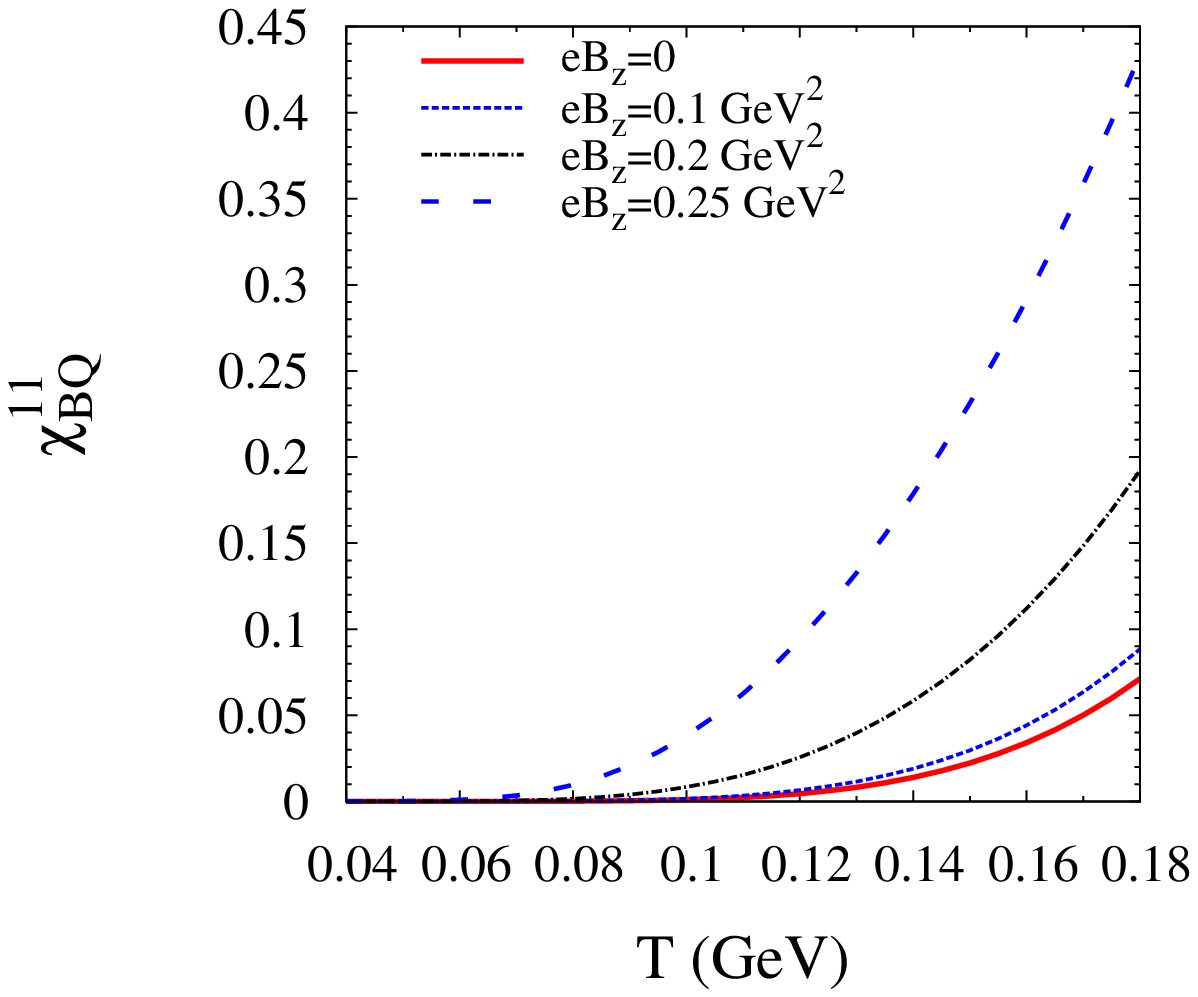}\label{fig:chi_BS_T}}
 \caption{ $\chi_{BS}$, $\chi_{QS}$, $\chi_{BQ}$ as a function of temperature at $\mu=0$.}
 \label{fig:correlation}
\end{figure*}

Figure~\ref{fig:chi_T} shows the variation of second order ($\chi^2$) and fourth order 
($\chi^4$) susceptibilities of different conserved charges with  $T$ at $\mu=0$.
At low $T$, dominating contribution to $\chi_B$ comes from protons and neutrons. With
increase in $T$, other heavier baryons populate and contribute to $\chi_B$ and hence 
susceptibilities increase with increase in temperature.  Since all the baryons have baryon 
number $\pm 1$, there is no difference in magnitude for higher order susceptibility for 
$\chi_B$. Both $\chi_B^2$ and $\chi_B^4$ increase with increase in $B$ especially at high 
temperature.  The number density of protons increase with the magnetic field. At high 
temperature the $\Delta$ particles get excited. The number density of $\Delta^{++}(1232)$ particles 
increase with magnetic field even faster than that for the proton for two reasons.   Firstly, 
 as the spin of $\Delta^{++}(1232)$ 
 is $3/2$  the effective mass of $\Delta^{++}(1232)$ decreases with magnetic field~\cite{Endrodi_HRG}. Secondly, for a spin 
 $3/2$ and doubly charged particle the effect of degeneracy and magnetic field make the 
 density increase faster. As a result the susceptibilities increase perceptibly at high temperature 
 with the increase in magnetic field.

The dominant contribution to $\chi_S$  at low temperatures come from kaons 
which have strange quantum number $\pm 1$. As a result magnitudes of $\chi_S^2$ and $\chi_S^4$ are similar
in low $T$ region up to  $0.1$ GeV.
Since $K^{\pm}$ are spin zero particles, their populations get suppressed in presence of $B$. 
However, since $T$ is small, population itself is small and hence this suppression is also small. 
With increase in $T$, other strange mesons, like $K^*$ (spin one), starts populating the system. In presence
of magnetic field populations of $K^{*\pm}$ increases.  As the  contributions of $K^{\pm}$ and $K^{*\pm}$ to the susceptibilities are 
opposite in nature, in presence of $B$, up to certain temperature ($T < 0.12 GeV $), 
susceptibilities for strange quantum number do not get affected much by the magnetic field. At higher temperatures, 
other charged strange hadrons (like $\Sigma^{\pm}$) start populating the system. Furthermore, at 
high temperature and high magnetic field, the contribution of $K^{*\pm}$, 
to the susceptibilities, is much more compared to that for $K^{\pm}$. This also makes the susceptibilities increase 
with magnetic field. 
Also, at high $T$, several strange
baryons like $\Xi, \Omega$ are excited which have strange quantum number
$\pm2$ and $\pm3$ respectively. Therefore, in the high $T$ region, magnitude of $\chi_S$ increase for higher orders. 

In the low temperature domain, 
$\chi_Q^2$ and $\chi_Q^4$ decrease in presence of magnetic field. Beyond certain temperature, 
those increase with increase in magnetic field. Also this limiting temperature
decreases with increase in magnetic field. 
The dominant contribution to the susceptibilities of conserved electric charges at low temperatures 
come from pions which are spin zero particles. 
Next dominating contributions are coming from kaons which are also spin zero particles.
Therefore, the populations of these particles 
get suppressed in presence of magnetic field.
As a result susceptibilities of conserved electric charges
decrease with increase of magnetic field at low temperature. 
With increase of temperature  $\rho$, $ K^*$, $p$, $\Delta$ etc., which are spin non-zero particles, appear in the system and these 
particles cause the susceptibilities to rise at high temperature.

\begin{figure*}[!htb]
\vspace{-0.1in}
\centering
{\includegraphics[scale=0.39]{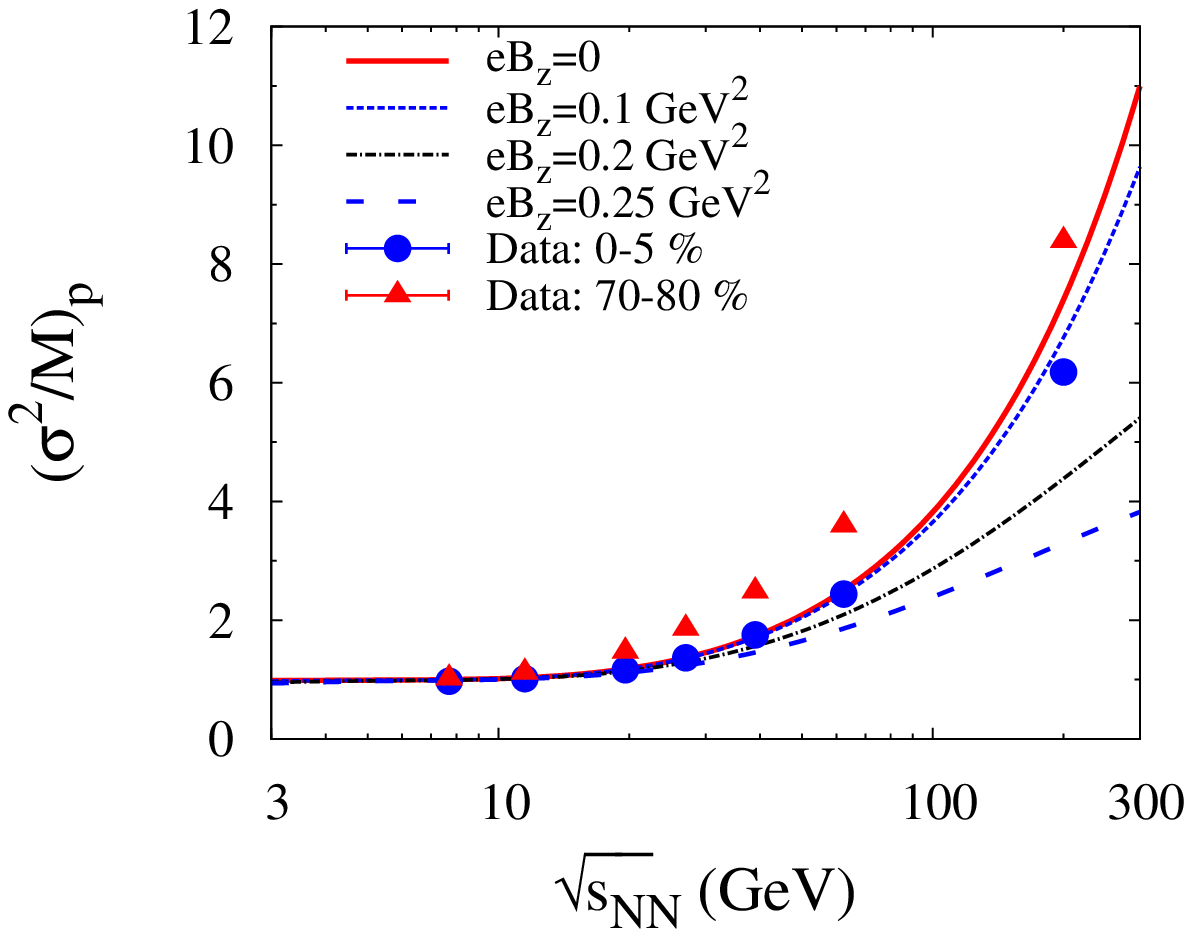}\label{fig:chi_2_1_np_roots}}
{\includegraphics[scale=0.39]{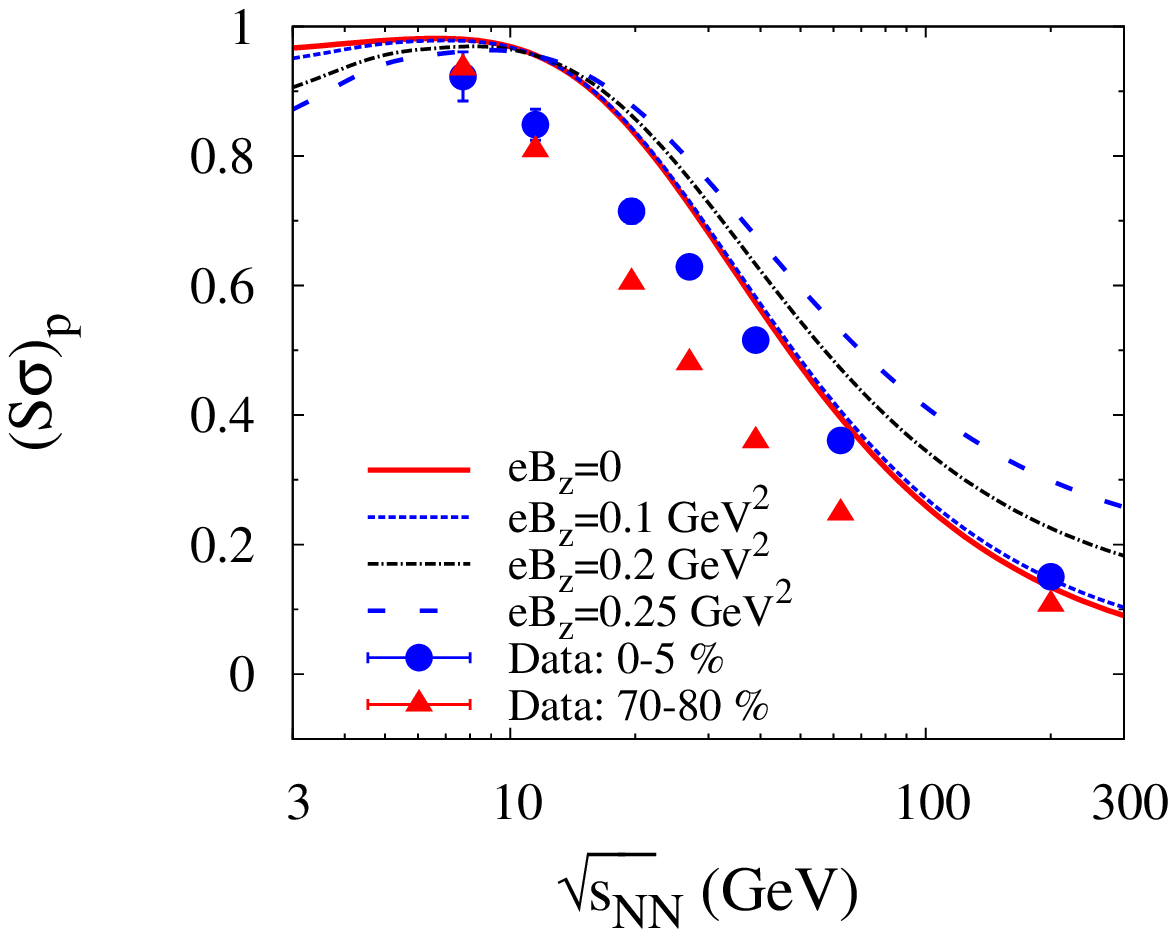}\label{fig:chi_3_2_np_roots}}
{\includegraphics[scale=0.39]{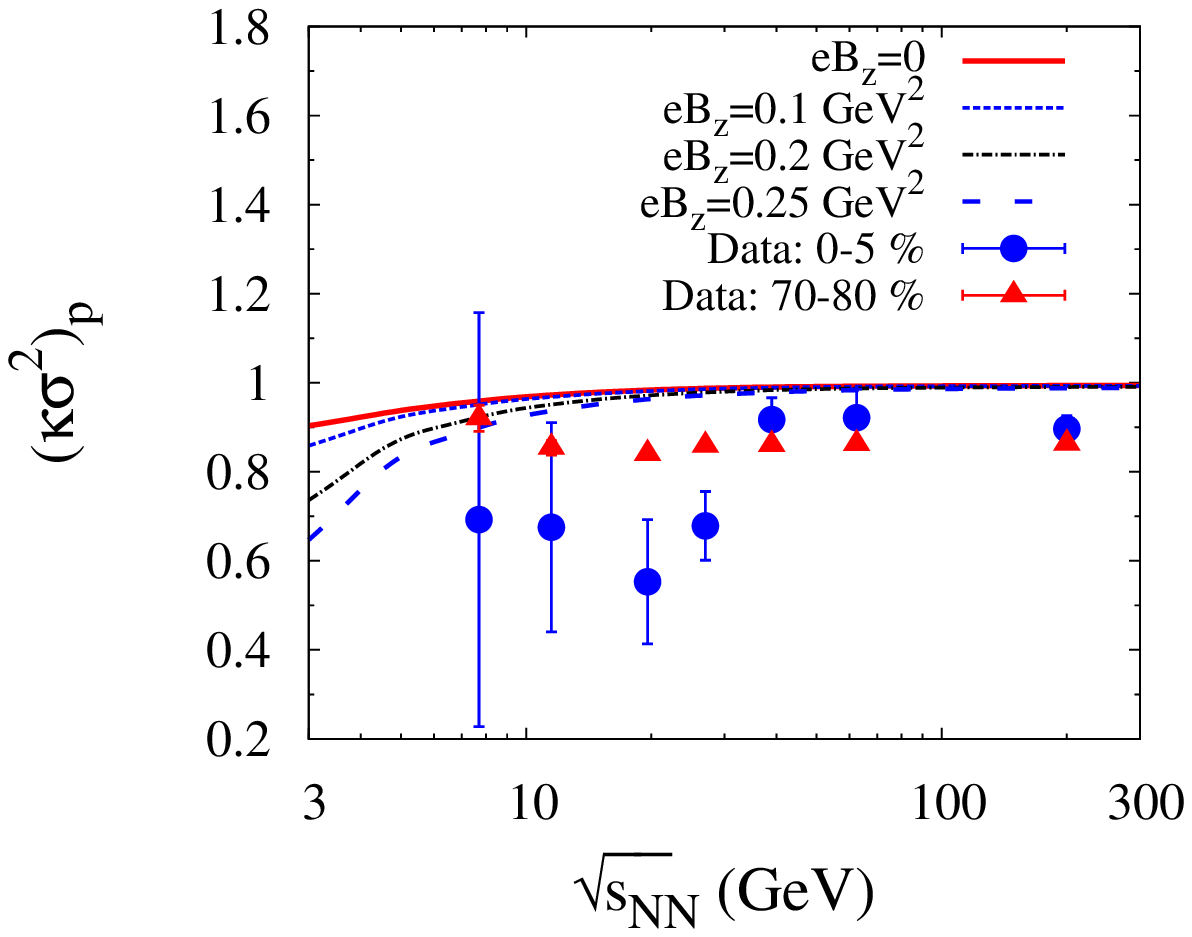}\label{fig:chi_4_2_np_roots}}
\vspace{-0.1in}
{\includegraphics[scale=0.39]{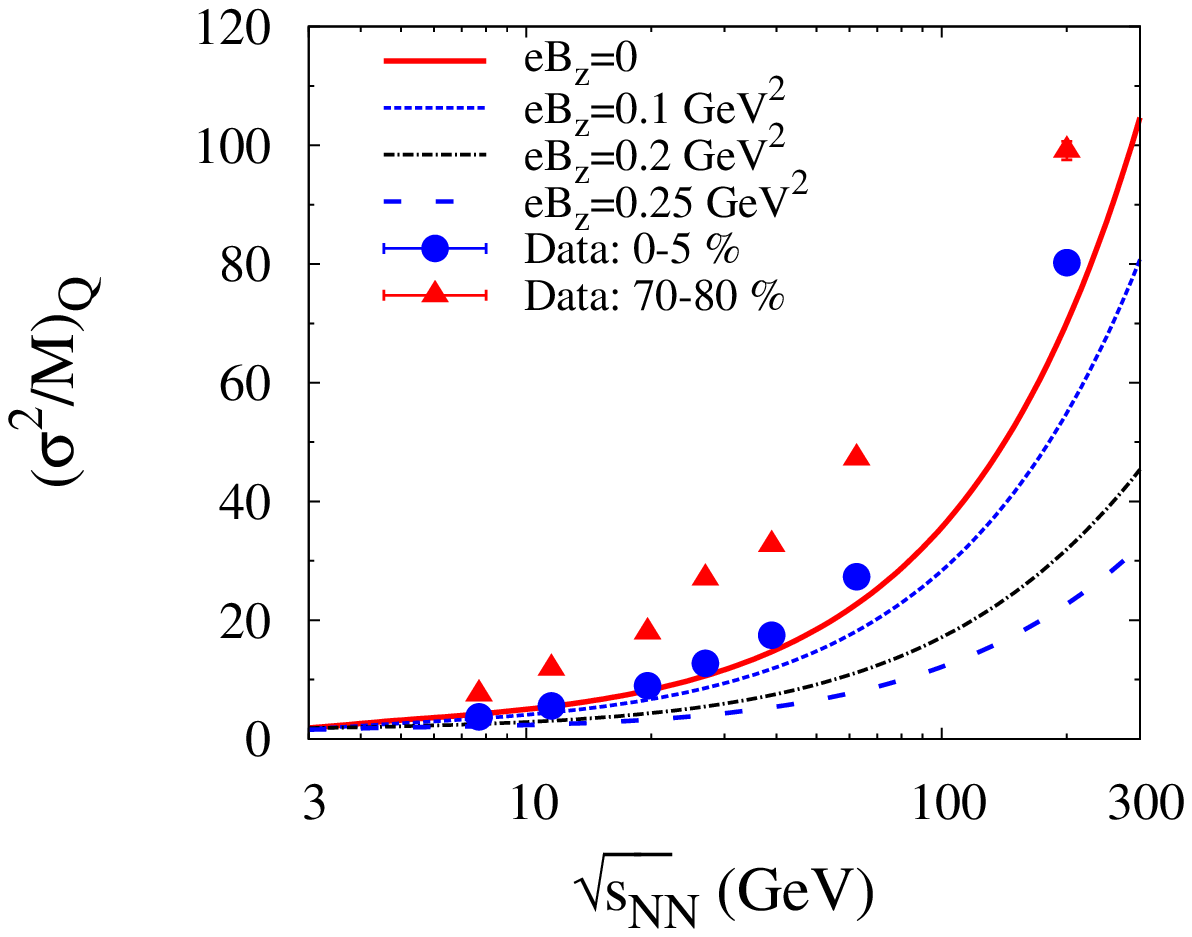}\label{fig:chi_2_1_nc_roots}}
{\includegraphics[scale=0.39]{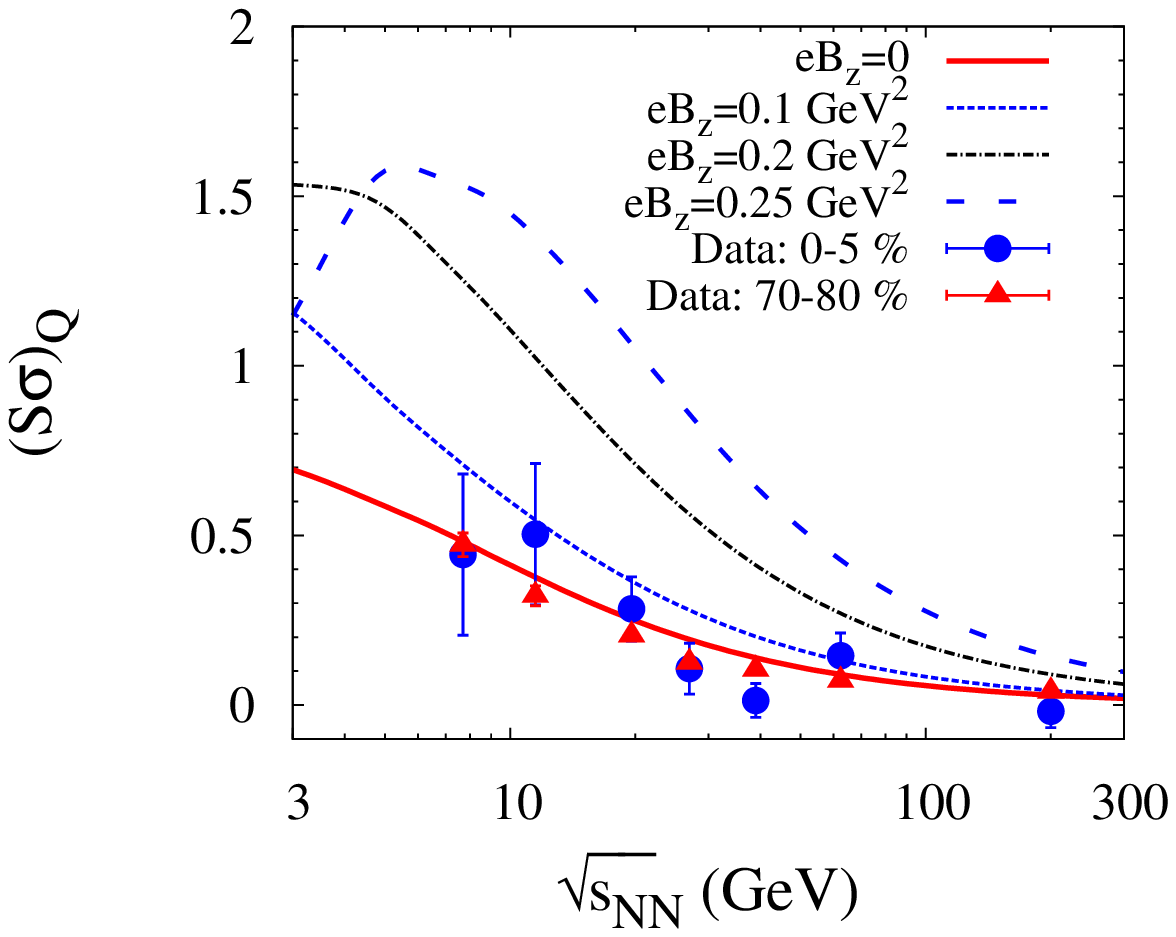}\label{fig:chi_3_2_nc_roots}}
{\includegraphics[scale=0.39]{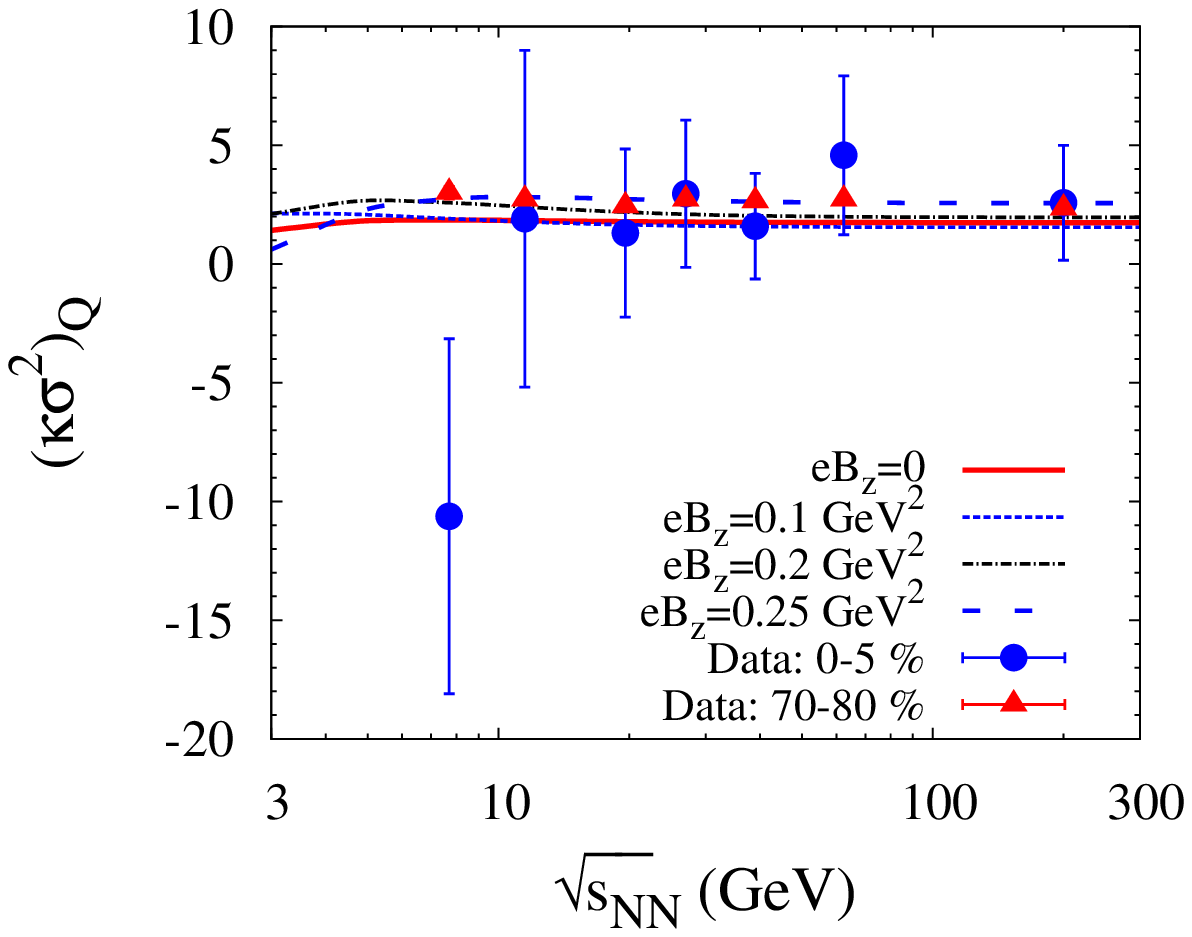}\label{fig:chi_4_2_nc_roots}}
\vspace{-0.1in}
{\includegraphics[scale=0.39]{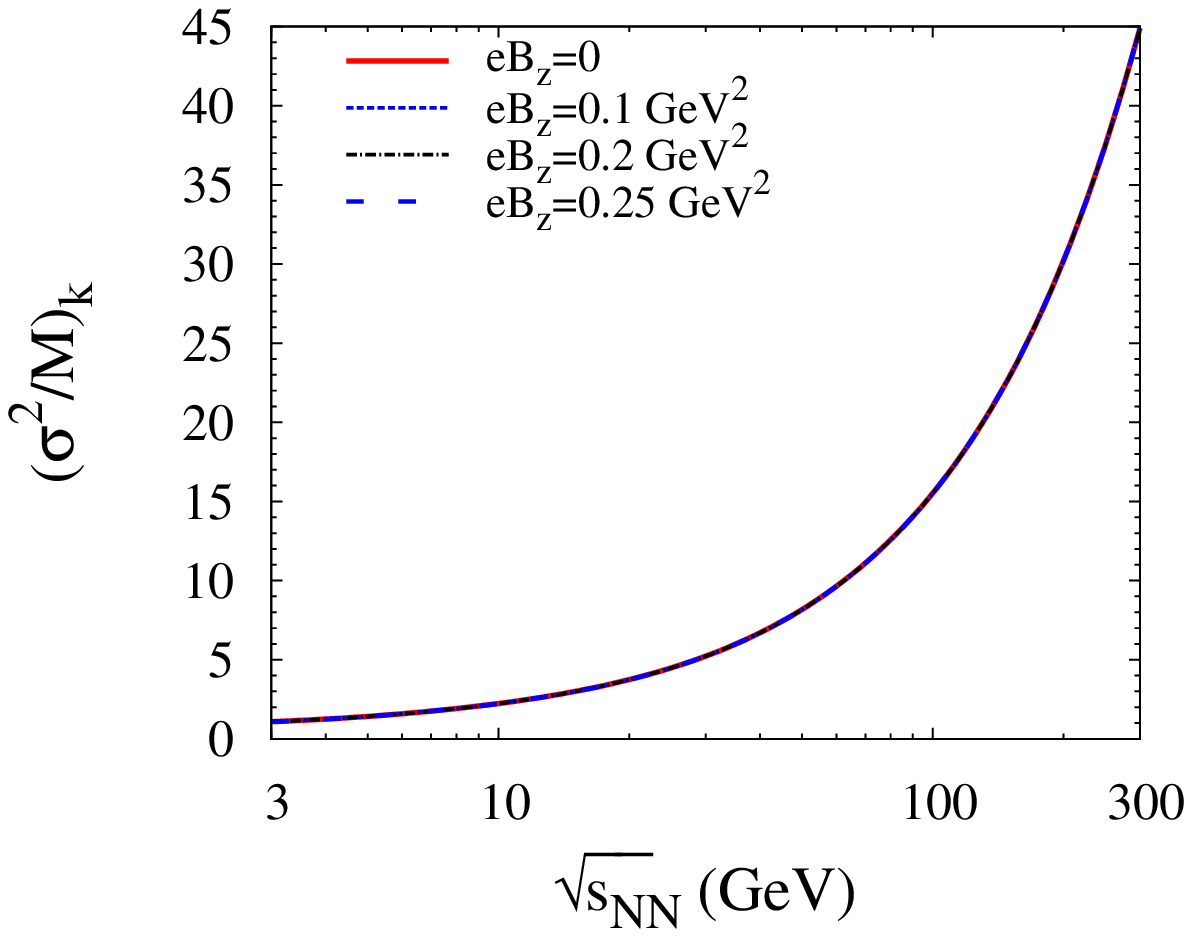}\label{fig:chi_2_1_nk_roots}}
{\includegraphics[scale=0.39]{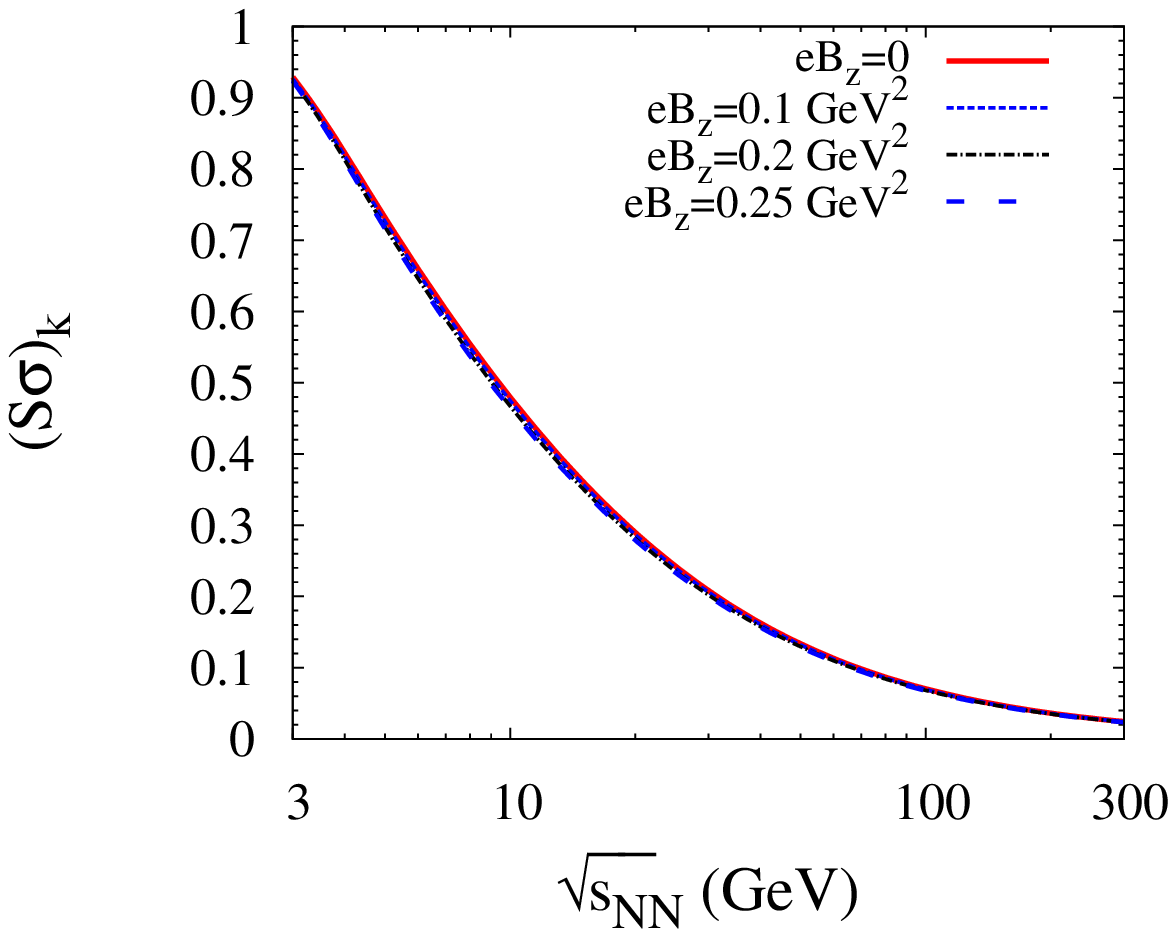}\label{fig:chi_3_2_nk_roots}}
{\includegraphics[scale=0.39]{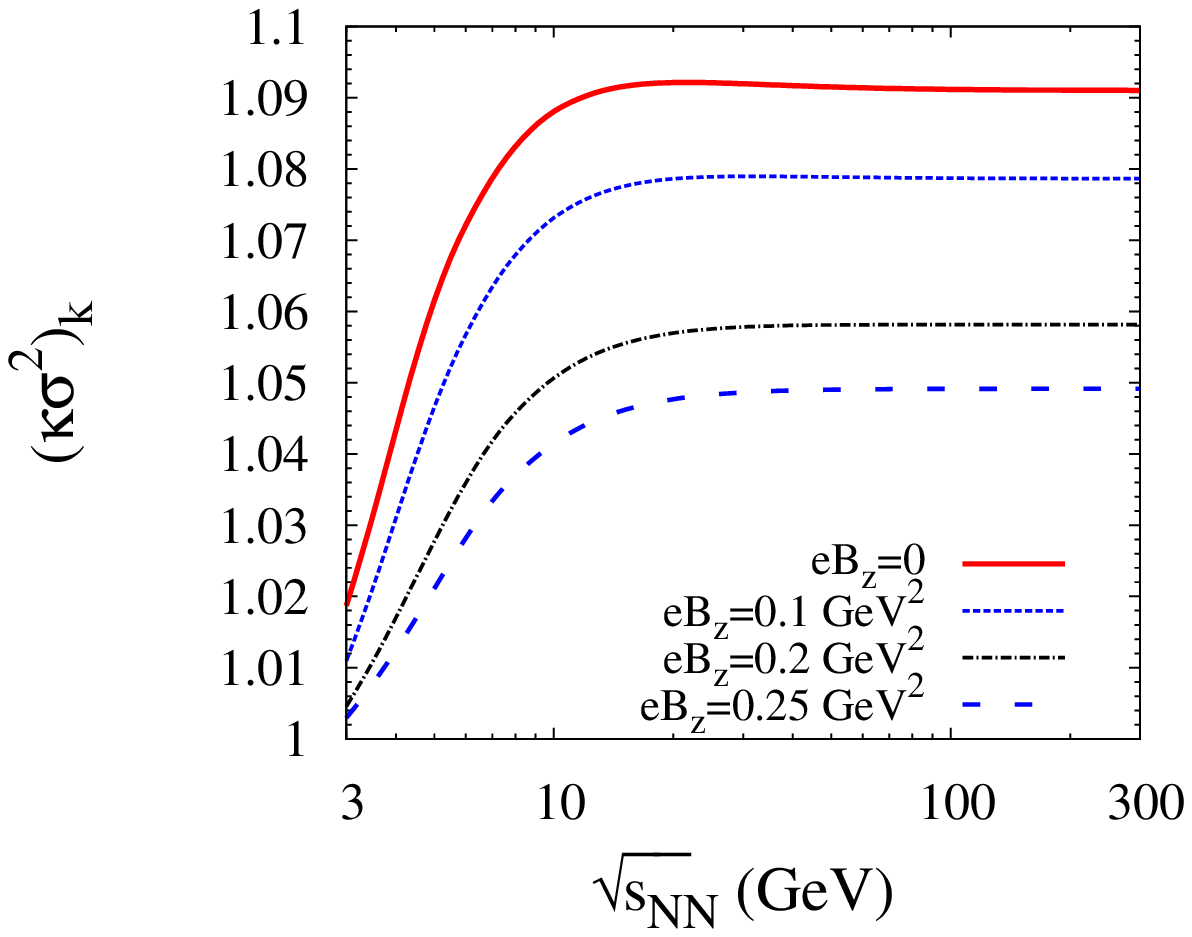}\label{fig:chi_4_2_nk_roots}}
\caption{Products of moments  of net-proton, net-charge and net-kaon as  functions of centre of mass energy.}  
\label{fig:chi_np_roots}
\end{figure*}

 In figure~\ref{fig:correlation} we have plotted the correlations (off diagonal susceptibilities) of 
 conserved charges.   For  all the cases the correlations increase with  magnetic field. The 
 maximum effect is observed in the baryon charge sector as the baryons with higher spin states 
 and charge contribute in this sector. Here the leading contributors are $p$, $\Sigma^\pm$ and 
 $\Delta$. For all these particles the number density increases with increase in magnetic field, 
 especially at high temperature, as a result the correlation increases. The leading order contribution 
 to $BS$ comes from $\Lambda$ and
$\Sigma^\pm$. As the number densities of these particles 
increase in presence of a magnetic field the correlation also increases, especially at high 
temperature. For the $QS$ sector, the correlation is suppressed at low temperature, in presence of 
a magnetic field,  as the major contributions come from $K^\pm$. The number density of $K^\pm$, being a spin $0$ particle, decreases 
 in presence of a magnetic field. As the temperature increases the spin $1$ particle $K^{*\pm}$ appear in the medium. 
 This makes the $QS$ correlation increase at high temperature.

\subsection{Beam energy dependence of the products of moments}

Experimentally measured moments such as mean $(M )$, standard deviation
$(\sigma)$, skewness $(S)$ and kurtosis $(\kappa)$ of conserved charges
are used to characterize the shape of charge distribution.  The products
of moments can be linked with susceptibilities by the following
relations,
\begin{equation}\label{moment_product}
\frac{\chi_x^2}{\chi_x^1}=\frac{\sigma_x^2}{M_x}, ~~~~
\frac{\chi_x^3}{\chi_x^2}=S_x\sigma_x, ~~~~
\frac{\chi_x^4}{\chi_x^2}=\kappa_x\sigma_x^2.
\end{equation}
These ratios are independent of the volume of the system and  play
crucial role for the search of possible critical point in the QCD phase
diagram. To make contact between our model and experimental data we need
a parametrization of $T$, $\mu$'s and $B$ with $\sqrt{s}$. Such
parametrizaion exist for multiplicities of identified hadrons in the
absence of magnetic field~\cite{Cleymans_PRC_73,PLB695_Karsch} for
central collisions. Here we use this parametrization to check effects of
magnetic fields on the fluctuation ratios of net proton and net charge
data given in~\cite{STAR_PRL_112,STAR_arXiv:1402.1558}.  However with
this parametrization, a one-to-one comparison with experimental data is
not expected to be perfect as the system may not have equilibrated
completely, leading to differences of fluctuations from those at thermal
and chemical equilibrium. Secondly the parameters for central collisions
are not same as peripheral collisions. Lastly the effects of magnetic
field has not been included in the above parametrization. We therefore
seek a qualitative comparison of the fluctuations with non-zero magnetic
fields vis-a-vis those at $B=0$. A further example for possible future
net-kaon fluctuation ratios is also presented.

Figure~\ref{fig:chi_np_roots} shows the of products of moments for
net-proton, net-charge and net-kaon as functions of centre of mass
energy $\sqrt{s}$. The two ratios $(\sigma^2/M)_x$ and $(S\sigma)_x$
involve odd derivatives $\chi_1$ and $\chi_3$ respectively. These should
ideally vanish at zero densities. On the other hand, for large densities
these approach 1. On the other hand, the ratio $(\kappa \sigma^2)_x$ is
almost always close to 1 as obtained in HRG. This defines the general
behaviour of these ratios as functions of $\sqrt{s}$.

With non-zero magnetic field We find that $(\sigma^2/M)_p$ is
very sensitive with increasing $\sqrt{s}$. For $(S\sigma)_p$ the
magnetic field effect is non-monotonic with $\sqrt{s}$. Both these
behaviours are prominently effects from the anomalous magnetic moment of
the proton. We checked that without this anomalous part these ratios
have insignificant dependence on the magnetic field. For $(\kappa
\sigma^2)_p$ there is a small magnetic field dependence only at very low
$\sqrt{s}$.  For net-charge we find $(\sigma^2/M)_Q$ to have similar
dependence on magnetic field as that of $(\sigma^2/M)_p$, but the
magnitude is now larger due to contributions of the low mass charged
mesons. $(S\sigma)_Q$ is found to be most sensitive to the magnetic
field for lower $\sqrt{s}$. In these cases the effect of anomalous
magnetic moment is subdominant. The dependence of $(\kappa\sigma^2)_Q$
on magnetic field is small and of the same order as corresponding ratio
of net charge.

By construction, the 0-5$\%$ centrality data is close to the HRG results
for $B=0$. However the sensitivity of $B$ underlines its importance to
be considered as a freeze-out parameter in future. On one hand it seems
that most magnetic field effects would be observed for peripheral
heavy-ion collisions because large number of spectator particles would
give rise to large magnetic field. On the other hand a larger system
created in the central collisions may sustain even somewhat smaller
magnetic field for a considerable time~\cite{Tuchin_PRC_82}.  In general
it seems that the magnetic field affects the lower order fluctuation
ratios the most. Therefore $(\sigma^2/M)_x$ are suitable observables for
fixing the magnetic field value in the HRG model.

Finally we present a case for net-kaons. As we saw that the strangeness
fluctuations are weakly dependent on the magnetic field, so is the case
for all the ratios ($\sigma^2/M)_k$, $(S\sigma)_k$
and $(\kappa \sigma^2)_k$. So strangeness fluctuations are not much
useful for detection of effects from magnetic field. We note here that
no anomalous magnetic moments were introduced for the mesons as no
conclusive experimental data exist for them.

\section{Conclusion}\label{secconclusion}

We have studied the fluctuations of conserved charges, namely baryon,
strangeness and electric charge, using HRG model in presence of a
magnetic field. We have chosen three values of magnetic fields, highest
of which is close to the predicted magnetic field for non-central HIC at
LHC energy.  The study was done with several assumptions like neglecting
interactions among the hadrons, ignoring the mass and size modifications
in presence of temperature, chemical potential and magnetic field, as
well as assuming that the effect of magnetic fields survive long enough
for the hadrons to get affected by it. This study basically sets an
upper limit for the magnetic field effects. 

The baryon number and electric charge number susceptibilities are more
sensitive to the magnetic field as compared to the strangeness number
susceptibilities. Among the correlators, the baryon$-$charge correlation
is found to be the most sensitive.  The ratios of fluctuations that are
measurable experimentally also manifest some effects of the magnetic
field. We loosely chose thermodynamic parameters fitted for
multiplicity at central collisions for $B=0$ to see the deviation in an
external magnetic field. We found that the fluctuation ratios for the
net kaon have a soft dependence on the magnetic field whereas for net
proton and net electric charge the dependence is quite strong. A
reparametrization of freeze-out conditions with $B$ as a parameter may
be in order and will be studied elsewhere.

\section{Acknowledgement}
We thank CSIR, DST and AvH foundation for support. SS thanks S. Das, S. Prasad, 
M. Younus and S. Maity for discussion.

\end{document}